\documentclass[aps,preprint,preprintnumbers,nofootinbib,showpacs,superscriptaddress]{revtex4-1}
\usepackage{graphicx,color,amsmath}
\begin{document}
\title{Rare $B^-\to \Lambda\bar p \nu \bar\nu$ decay}

\author{C.Q. Geng}
\affiliation{Department of Physics, National Tsing Hua University, Hsinchu, Taiwan 300, R.O.C.}
\affiliation{Physics Division, National Center for Theoretical Sciences, Hsinchu, Taiwan 300, R.O.C.}
\author{Y.K. Hsiao}
\affiliation{Physics Division, National Center for Theoretical Sciences, Hsinchu, Taiwan 300, R.O.C.}
\date{\today}
\begin{abstract}
We study the four-body semileptonic baryonic decay of $B^-\to \Lambda\bar p\nu\bar \nu$
in the standard model. We find that the decay branching ratio is $(7.9\pm 1.9)\times 10^{-7}$.
Similar to the rare decays of $B^-\to K^{(*)-}\nu\bar \nu$, this baryonic decay of $B^-\to \Lambda\bar p\nu\bar \nu$
is also sensitive to new physics and accessible to the future $B$ factories.
\end{abstract}


\maketitle
\newpage
\section{introduction}
Since the inclusive flavor-changing neutral-current (FCNC) processes of $b\to s\nu\bar \nu$ and $b\to d\nu\bar \nu$
in the standard model (SM) can only be induced through box and electroweak penguin diagrams,
the corresponding exclusive ones, such as $B\to K^{(*)}\ell\bar \ell$ ($\ell=e,\mu,\tau, \nu$)  are highly suppressed as rare decays.
Experimental searches for these rare decays could shed light to find new physics.
For example, as the current experimental upper bound on
 the decay branching ratio of $B^-\to K^-\nu\bar \nu$
  is $14\times 10^{-6}$~\cite{BexMnunu1,BexMnunu2}, while its SM prediction is
 $(4.5\pm 0.7)\times 10^{-6}$~\cite{BthMnunu1}, there would exist some kind of new physics
\cite{BthMnunu1,ED,UP,IS,4th} between the sandwiched area.
Moreover, for  the decay of $B\to  K^{*}\nu\bar \nu$ with an on-shell $ K^{*}\to K\pi$,
some physical   observables~\cite{ChenGeng} are also available to test T-violating effects beyond the SM.
When the experimental sensitivities are gradually improved,
even the decays of $B^-\to(\pi^-,\rho^-)\nu\bar \nu$ via $b\to d\nu\bar \nu$ with an additional suppression of $|V_{td}/V_{ts}|^2$
can also function as probes for new physics.

In this report, we propose to use the baryonic modes of $B^-\to \Lambda\bar p\ell\bar \ell$ ($\ell=e,\mu,\tau, \nu$)
as a new type of the exclusive $B$ decays via $b\to s\ell\bar \ell$ to examine FCNCs.
  To simplify our discussions on the baryonic form factors, we will concentrate on the four-body semileptonic baryonic decay of
$B^-\to \Lambda\bar p\nu\bar \nu$.
In particular, we will study its decay branching ratio in the SM.
It is interesting to note that the $B^-\to \Lambda\bar p\nu\bar \nu$  can be well reconstructed
experimentally 
since 
 the charged $\bar p$ along with $p \pi^-$ from $\Lambda$  can be easily detected.

The decay of  $B^-\to \Lambda\bar p\nu\bar \nu$ has  several interesting features.
First, as a four-body decay, the observables for angular distribution asymmetries
can be constructed as a probe to
right-handed vector as well as  (pseudo-)scalar currents beyond the SM.
Second, by keeping the $\Lambda$ spin $\vec{s}_\Lambda$, we are able to construct a T-odd observable
$\vec{s}_\Lambda\cdot (\vec{p}_\Lambda\times \vec{p}_{\bar p})$ with
the $\Lambda(\bar p)$ momentum $\vec{p}_{\Lambda(\bar p)}$
to test time reversal violation.
As the basis to study new physics,  ${\cal B}(B^-\to \Lambda\bar p\nu\bar \nu)$ in the SM
can be naively estimated to be of order $10^{-6}-10^{-7}$.
This is in comparison with the $B^-\to p\bar p \ell^-\bar \nu$ via $b\to u\ell^-\bar \nu$
being 100 times bigger than $b\to s\nu\bar\nu$,
while the predicted ${\cal B}(B^-\to p\bar p \ell^-\bar \nu)$ is of order $10^{-4}$ to $10^{-5}$ \cite{ppenu}.
In order to precisely calculate the decay, a knowledge of
the matrix elements for the $B^-\to \Lambda\bar p$ transition is needed,
which is difficult to obtain in QCD.
However, since $B^-\to \Lambda\bar p\nu\bar \nu$ is considered to
associate with the three-body baryonic $\bar B$ decays
of $\bar B\to p\bar p \,(\bar K^{(*)},\,\pi,\,\rho)$
\cite{ppKLambdapbarpi_Belle,ppK(star)pi_Belle,ppK_Babar,ppKpi_Belle,pppi_Babar,ppKstar_Belle}
and $\bar B^0\to p\bar p D^{(*)0}$ \cite{{ppD(star)_Belle,ppD(star)_Babar}}
via the $\bar B\to {\bf B\bar B'}$ transitions,
the solution can be simply made.
The parameterizations for
the $\bar B\to {\bf B\bar B'}$ transitions
in \cite{Hou2,GengHsiaoHY,Hsiao,GengHsiao2,GengHsiao3,GengHsiao4,GengHsiao5} can be reliably adopted,
as the theoretical studies of
${\cal B}(\bar B\to \Lambda\bar \Lambda \bar K)$ \cite{GengHsiao2},
${\cal B}(\bar B^0\to \Lambda\bar \Lambda D^0)$ and ${\cal B}(B^-\to \Lambda\bar p D^{(*)0})$ \cite{GengHsiaoHY}
relating the $\bar B\to {\bf B\bar B'}$ transitions are approved
to agree with the data \cite{LambdaLambdaD,LambdapbarD}.
In addition, with the $B^-\to p\bar p$ transition,
the $CP$ violation for $ B^-\to p\bar p K^{*-}$ \cite{GengHsiao4} is found to be
nearly 20\% of the world average \cite{ppKstar_BABAR,ppKstar_BELLE}
even though it is still inconclusive experimentally due to the data errors.

The paper is organized as follows.
In  Sec. II, we provide the formalism, which involves
the decay amplitude and rate of $B^-\to \Lambda\bar p\nu\bar{\nu}$ based on the form factors in
the parameterizations for the matrix elements of
the $\bar B\to {\bf B\bar B'}$ transitions.
We give our numerical results and discussions  in Sec. III.
In Sec. IV, we present the conclusions.

\section{Formalism}
The effective Hamiltonian for the inclusive mode of $b\to s\nu_\ell \bar{\nu}_\ell$  is given by \cite{btosnunu}
\begin{eqnarray}\label{effH}
{\cal H}(b\to s \nu_\ell \bar{\nu}_\ell)=
\frac{G_F}{\sqrt 2}\frac{\alpha_{em}}{2\pi \text{sin}^2\theta_W }
\lambda_t D(x_t) \bar s\gamma_\mu(1-\gamma_5)b\bar{\nu}_\ell\gamma^\mu(1-\gamma_5)\nu_\ell\,,
\end{eqnarray}
with $\lambda_t=V_{ts}^*V_{tb}$, $x_t\equiv m_t^2/m_W^2$, and $\nu_\ell=\nu_e$ or $\nu_\mu$ or $\nu_\tau$,
where
$D(x_t)$ is the top-quark loop function \cite{topfn1,topfn2}.
\begin{figure}[t!]
\centering
\includegraphics[width=2.2in]{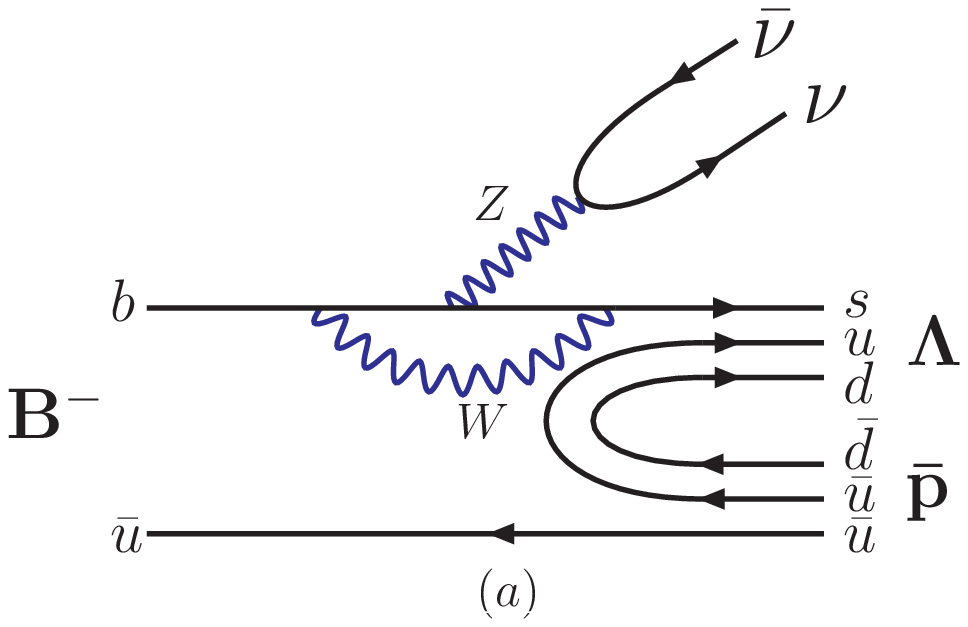}
\includegraphics[width=2.4in]{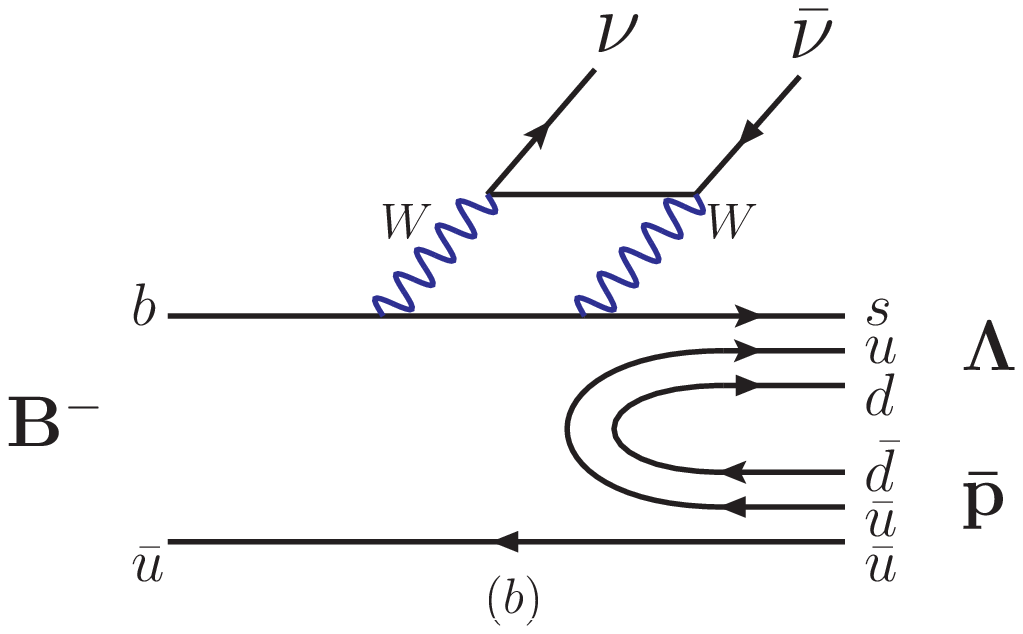}
\caption{Contributions to the $B^-\to \Lambda\bar p\nu\bar \nu$ decay from
(a)  penguin  and (b)  box diagrams. }\label{fig1}
\end{figure}
 From Fig. \ref{fig1}, via the effective Hamiltonian in Eq. (\ref{effH})
the amplitude of $B^-\to \Lambda\bar p\,\nu_\ell\bar \nu_\ell$
 can be factorized as
\begin{eqnarray}\label{amp}
{\cal A}(B^-\to \Lambda\bar p \nu_\ell\bar \nu_\ell)=
\frac{G_F}{\sqrt 2}\frac{\alpha_{em}}{2\pi \text{sin}^2\theta_W }\lambda_t D(x_t)
\langle \Lambda\bar p|\bar s\gamma_\mu (1-\gamma_5)b|B^- \rangle
\;\bar \nu_\ell\gamma^\mu (1-\gamma_5) \nu_\ell\;,
\end{eqnarray}
where the explicit form of the matrix element for $B^-\to \Lambda\bar p$ depends on the parameterization,
which has been studied in three-body baryonic $\bar B$ decays. With Lorentz invariance,
the most general forms of the $\bar B\to {\bf B\bar B'}$ transition form factors
are given by \cite{GengHsiaoHY}
\begin{eqnarray}\label{transitionF}
\langle {\bf B}{\bf\bar B'}|\bar q'\gamma_\mu b|\bar B\rangle&=&
i\bar u(p_{\bf B})[  g_1\gamma_{\mu}+g_2i\sigma_{\mu\nu}p^\nu +g_3p_{\mu} +g_4q_\mu +g_5(p_{\bf\bar B'}-p_{\bf B})_\mu]\gamma_5v(p_{\bf \bar B'}),~~\nonumber\\
\langle {\bf B}{\bf\bar B'}|\bar q'\gamma_\mu\gamma_5 b|\bar B\rangle&=&
i\bar u(p_{\bf B})[ f_1\gamma_{\mu}+f_2i\sigma_{\mu\nu}p^\nu +f_3p_{\mu} +f_4q_\mu +f_5(p_{\bf\bar B'}-p_{\bf B})_\mu]v(p_{\bf \bar B'}),~~~
\end{eqnarray}
with $q=p_{\bf B}+p_{\bf\bar B'}$ and $p=p_{\bar B}-q$,
for the vector and axial-vector quark currents, respectively.
For the momentum dependences,
the form factors $f_i$ and $g_i$ ($i=1,2, ...,5$) are taken to be \cite{GengHsiao3}
\begin{eqnarray}\label{figi}
f_i=\frac{D_{f_i}}{t^3}\,,\;\;g_i=\frac{D_{g_i}}{t^3}\,,
\end{eqnarray}
with $t\equiv q^2\equiv m_{\bf B\bar B'}^2$, where $D_{f_i}$ and $D_{g_i}$ are constants
to be determined by the measured data in $\bar B\to p\bar p M$ decays.
Note that $1/t^3$ arises from 3 hard gluons
as the propagators to form a baryon pair in the approach of the pQCD counting rules
\cite{Brodsky1,Brodsky2,Brodsky3,Hou2},
where two of them attach to valence quarks in ${\bf B}{\bf\bar B'}$,
while the third one kicks and speeds up the spectator quark in $\bar B$.
It is worth to note that, due to $f_i,\,g_i\propto 1/t^3$
 the dibaryon invariant mass spectrum
peaks at the threshold area and flattens out at the large energy region.
Hence, this so-called threshold effect measured as a common feature in $\bar B\to p\bar p M$ decays
should also appear in the $B^-\to \Lambda\bar p\nu_\ell\bar{\nu}_\ell$ decay.
To integrate over the phase space for the amplitude squared $|\bar {\cal A}|^2$,
which is obtained by assembling the required elements in Eqs. (\ref{amp}), (\ref{transitionF}), and (\ref{figi})
and summing over all fermion spins,
\begin{figure}[t!]
\centering
\includegraphics[width=2.8in]{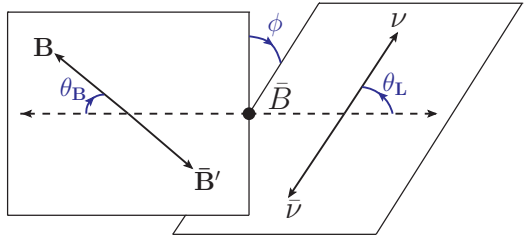}
\caption{Three angles $\theta_{\bf B}$, $\theta_{\bf L}$,
and $\phi$ in the phase space for the four-body $\bar B\to {\bf B\bar B'}\nu\bar \nu$ decay.
}\label{4body}
\end{figure}
the knowledge of the kinematics for the four-body decay
is needed. For this reason, we use the partial decay width~\cite{Kl4,Wise,Cheng:1993ah}:
\begin{eqnarray}\label{int1}
d\Gamma=\frac{|\bar {\cal A}|^2}{4(4\pi)^6 m_{\bar B}^3}X\beta_{\bf B}\beta_{\bf L}\, ds\, dt\, d\text{cos}\,\theta_{\bf B}\, d\text{cos}\,\theta_{\bf L}\, d\phi\,,
\end{eqnarray}
where
\begin{eqnarray}\label{int2}
X&=&\bigg[\frac{1}{4}(m_B^2-s-t)^2-st\bigg]^{1/2}\,,\nonumber\\
\beta_{\bf B}&=&\frac{1}{t}\lambda^{1/2}(t,m_{\bf B}^2,m_{\bf \bar B'}^2)\,,\nonumber\\
\beta_{\bf L}&=&\frac{1}{s}\lambda^{1/2}(s,m_{\nu}^2,m_{\bar \nu}^2)\,,
\end{eqnarray}
with $\lambda(a,b,c)=a^2+b^2+c^2-2ab-2bc-2ca$, and
$t$, $s\equiv (p_{\nu}+p_{\bar\nu})^2$, $\theta_{\bf B}$, $\theta_{\bf L}$, and $\phi$
are  five variables in the phase space.  As seen from Fig.~\ref{4body}, the angle
$\theta_{\bf B(L)}$ is between $\vec{p}_{\bf B}$  ($\vec{p}_{\nu}$)
in the $\bf B\bar B'$ ($\nu\bar \nu$) rest frame and
the line of flight of the $\bf B\bar B'$ ($\nu\bar \nu$) system in the rest frame
of the $\bar B$, while the angle $\phi$ is between the $\bf B\bar B'$ plane and  the $\nu\bar \nu$ plane,
which are defined by the momenta of the $\bf B\bar B'$ pair
and the momenta of the $\nu\bar \nu$ pair, respectively, in the rest frame of $\bar B$.
The ranges of the five variables are given by
\begin{eqnarray}\label{int3}
&&(m_\nu+m_{\bar \nu})^2\leq s\leq (m_{\bar B}-\sqrt{t})^2\,,\;\;
(m_{\bf B}+m_{\bf \bar B'})^2\leq t\leq (m_{\bar B}-m_\nu-m_{\bar \nu})^2\,,\nonumber\\
&&0\leq \theta_{\bf L},\,\theta_{\bf B}\leq \pi\,,\;\;0\leq \phi\leq 2\pi\,.
\end{eqnarray}
The decay branching ratio of ${\cal B}(B^-\to \Lambda\bar p \nu\bar \nu)$ depends on the integration in
Eqs. (\ref{int1}), (\ref{int2}) and  (\ref{int3}),
where we have to sum over the three neutrino flavors since they are indistinguishable.
We can also define the integrated angular distribution asymmetries, given by
\begin{eqnarray}
{\cal A}_{\theta_i}\equiv
\frac{\int^1_0 \frac{d{\cal B}}{dcos\theta_i}dcos\theta_i-\int^0_{-1}\frac{ d{\cal B}}{dcos\theta_i}dcos\theta_i}
{\int^1_0 \frac{d{\cal B}}{dcos\theta_i}dcos\theta_i+\int^0_{-1}\frac{ d{\cal B}}{dcos\theta_i}dcos\theta_i}\,,\ (i=B,\, L)\,.
\label{AS}
\end{eqnarray}

\section{numerical results and discussions}
\begin{figure}[b!]
\centering
\includegraphics[width=2in]{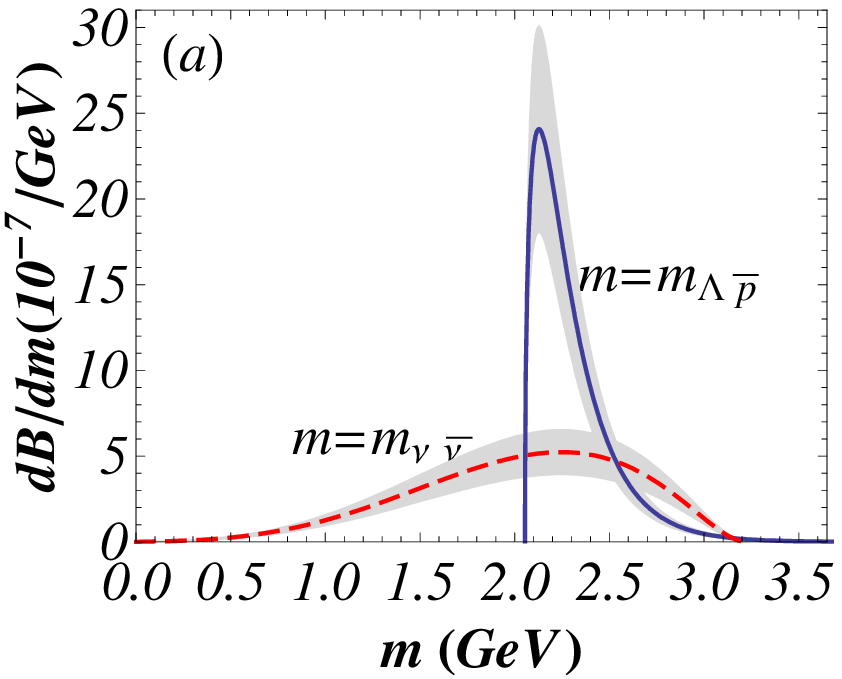}
\includegraphics[width=2.075in]{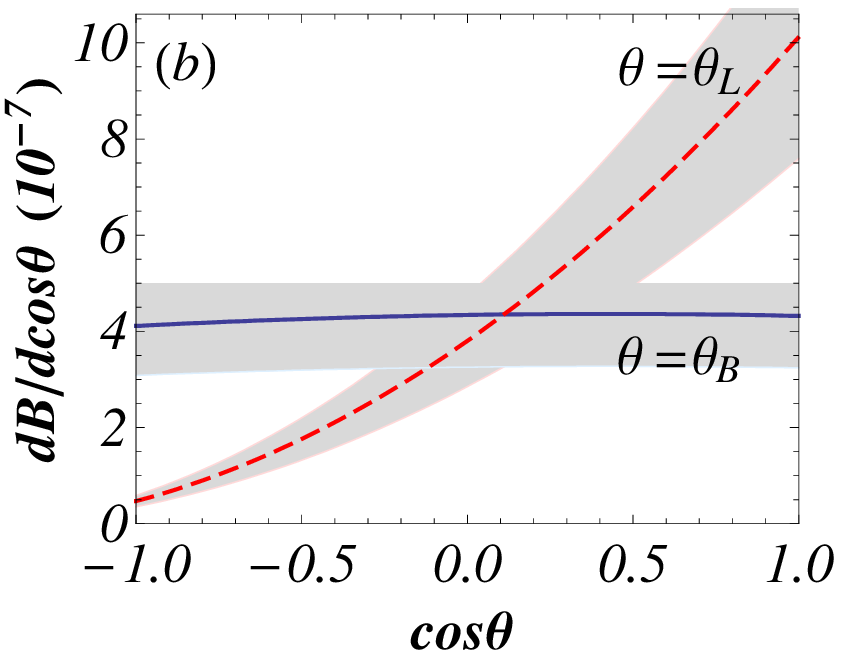}
\caption{Invariant mass spectra as functions of the invariant masses
$m_{\Lambda\bar p}$ and $m_{\nu\bar \nu}$ and angular distributions
as functions of $\text{cos}\theta_{B,L}$
for $B^-\to \Lambda\bar p \nu\bar \nu$, respectively,
where the shaded areas represent the theoretical uncertainties
from the form factors and CKM mixings.}\label{fig3}
\end{figure}

For the numerical analysis, we take  the values of $G_F$, $\alpha_{em}$,
$\text{sin}^2\theta_W$ and $V_{ts}^*V_{tb}$ in the PDG \cite{pdg} as the input parameters.
In the large t limit, the approach of the pQCD counting rules allows
the vector and axial-vector currents to be incorporated as two chiral currents.
As a result,
$D_{g_i}$ and $D_{f_i}$ from the vector currents can be related by
the another set of constants $D_{||}$ and  $D_{\overline{||}}$ from the chiral currents, and
the 10 constants for $B^-\to \Lambda\bar p$ are reduced as \cite{GengHsiaoHY}
\begin{eqnarray}\label{D||-Lambdapbar}
D_{g_1}=D_{f_1}=-\sqrt\frac{3}{2}D_{||}\,,\;D_{g_j}=-D_{f_j}=-\sqrt\frac{3}{2}D_{||}^j\,,
\end{eqnarray}
with $j=2,3, ..., 5$.
We note that the reduction is first developed in Refs. \cite{Brodsky1,Brodsky2,Brodsky3}
for the spacelike ${\bf B\to B'}$ baryonic form factors,
and extended to deal with the timelike $0\to {\bf B\bar B'}$ baryonic form factors
and the $\bar B\to {\bf B\bar B'}$ transition form factors
in the studies of the $\bar B\to {\bf B\bar B'}M$ decays
\cite{HY1,HY2,Hou1,Hou2,Hou3,Hsiao,GengHsiao5,GengHsiaoHY,GengHsiao1,GengHsiao2,GengHsiao3}.
 For $D_{||}^{(j)}$ and $D_{\overline{||}}^{(j)}$,
 we adopt the values, given by \cite{GengHsiaoHY}
\begin{eqnarray}\label{inputD}
&&(D_{||},\;D_{\overline{||}})=(67.7\pm 16.3,\,-280.0\pm 35.9)\;{\rm GeV^5},\nonumber\\
&&(D_{||}^2,\,D_{||}^3,\,D_{||}^4,\,D_{||}^5)=\nonumber\\
&&(-187.3\pm 26.6,\,-840.1\pm 132.1,\,-10.1\pm 10.8,\,-157.0\pm 27.1)\;{\rm GeV^4}\;,
\end{eqnarray}
extracted from the measured data of
the total branching ratios, invariant mass spectra, and angular distributions
in the $\bar B\to p\bar p M$ decays.
By using the various inputs,
we obtain the numerical results for the branching ratio
and angular distribution distribution asymmetries
of $B^-\to \Lambda\bar p \nu\bar \nu$ in Table \ref{tab1}, where
the values of $B^-\to p\bar p e^-\bar \nu_e$ are taken from Ref. \cite{ppenu}.
The invariant mass spectra and angular distributions
for $B^-\to \Lambda\bar p \nu\bar \nu$ are shown in Fig. \ref{fig3},
where the shaded areas represent the theoretical uncertainties
from the form factors and CKM mixings.
Note that the errors of the integrated angular asymmetries ${\cal A}_{\theta_{B,L}}$ in Table \ref{tab1}
are relatively small compared to those in Fig. \ref{fig3}b.
The reason is that ${\cal A}_{\theta_{B,L}}$ depend on
the ratios as shown in Eq. (\ref{AS}), which reduce the uncertainties.

\begin{table}[t!]
\caption{
Numerical results for $\cal B$ and ${\cal A}_{\theta_i}$ ($i=B,\,L$)
for $B^-\to \Lambda\bar p \nu\bar \nu$ and
$B^-\to p\bar p e^-\bar \nu_e$~\cite{ppenu}, respectively,
where the theoretical errors are mainly from the uncertainties in the form factors and CKM mixings.
}\label{tab1}
\begin{tabular}{|c|c|c|}
\hline
   & $B^-\to \Lambda\bar p \nu\bar \nu$&  $B^-\to p\bar p e^-\bar \nu_e$~\cite{ppenu}\\\hline
$\cal B$                &$(7.9\pm 1.9)\times 10^{-7}$&$(1.04\pm 0.29)\times 10^{-4}$\\
${\cal A}_{\theta_B}$   &$0.01\pm 0.02$ &$0.06\pm 0.02$\\
${\cal A}_{\theta_L}$   &$0.56\pm 0.02$ &$0.59\pm 0.02$\\
\hline
\end{tabular}
\end{table}

From Fig.~\ref{fig3}a, we see that ${\cal B}(B^-\to \Lambda\bar p \nu\bar \nu)$
receives the dominant contribution near the threshold of $m_{\Lambda\bar p}\to m_\Lambda+m_{\bar p}$,
when the curve sharply peaks in the invariant mass spectrum.
This reflects the fact of $1/t^3$ as the momentum dependence
in the $B^-\to \Lambda\bar p$ transition form factors.
In contrast, the curve in the $m_{\nu\bar\nu}$ spectrum is associated with 
the total energy of  the $\nu\bar \nu$ pair.
This is due to the helicity structure of $\bar \nu\gamma^\mu(1-\gamma_5)\nu$
in the amplitude,  formed as $(E_{\nu}+E_{\bar\nu})\varepsilon^\mu_-(p)$
with $\varepsilon^\mu_-(p)$ the left-handed polarization.
Moreover, the fact that
 $\varepsilon^\mu_-(p)$ couples to the left-handed helicity state of the virtual $Z$ boson
results in a factor of $(1+\text{cos}\theta_L)^2$ to
explain the angular distribution for $\theta=\theta_L$ in Fig. \ref{fig3}b.
As a duplicate case,  $B^-\to p\bar p e^-\bar \nu_e$ has the same helicity structure for the lepton pair
to couple to the left-handed helicity state of the virtual weak boson $W^{*-}$.
As a result,
it is  reasonable to have
${\cal A}_{\theta_L}(B^-\to \Lambda\bar p \nu\bar \nu)\simeq
{\cal A}_{\theta_L}(B^-\to p\bar p e^-\bar \nu_e)$ in Table \ref{tab1}.
On the other hand, since ${\cal B}(B^-\to \Lambda\bar p \nu\bar \nu)$
can be traced back to the tensor terms $f_2(g_2)$ in the $B^-\to\Lambda\bar p$ transition,
which give the main contributions,
$f_1\bar u\gamma_\mu\gamma_5\nu$ and $g_1\bar u\gamma_\mu\nu$ are too small to provide
factors of $(1\pm\text{cos}\theta_{\bf B})^2$ as apparent angular dependent terms, as given
in Fig.~\ref{fig3}b for $\theta=\theta_B$ and Table~\ref{tab1} for ${\cal A}_{\theta_B}$.

The domination of the tensor terms $f_2(g_2)$ in the $B^-\to\Lambda\bar p$ transition
can be realized.
The terms $f_3(g_3)$ disappear due to  $\varepsilon^\mu_-(p)$ with $p=p_\nu+p_{\bar \nu}$,
leading to the coupling of $\varepsilon_-\cdot p=0$.
Because of the relatively  small value of  $|D_{||}^4|\simeq 10\,\text{GeV}^4$, the terms $f_4(g_4)$ are negligible.
The suppression for $f_5(g_5)$ is in accordance with the limit of
$(p_{\bar p}-p_{\Lambda})_\mu=(E_{\bar p}-E_\Lambda,\,\vec{p}_{\bar p}-\vec{p}_{\Lambda})\to (0,\,\vec{0})$
as the invariant mass $m_{\Lambda\bar p}$ approaches the threshold area
to receive the main contribution for  ${\cal B}(B^-\to \Lambda\bar p \nu\bar \nu)$
(see Fig. \ref{fig3}a).
Moreover, with an additional $p$ in $f_2(g_2)\sigma_{\mu\nu}p^\nu$,
the ratio of $|f_2(g_2)p|^2$ to $|f_1(g_1)|^2$, which is equal to $D_{f_2(g_2)}^2|p|^2/D_{f_1(g_2)}^2\simeq 8|p|^2$,
can be enhanced by $|p|\to m_{\bar B}-(m_\Lambda+m_{\bar p})$ around the threshold area.
These explain
why $f_2(g_2)$ prevail over the other terms
in the $B^-\to \Lambda\bar p \nu\bar \nu$ decay.
Since the decays of $B^-\to p\bar p e^-\bar \nu_e$ and $B^-\to \Lambda\bar p \nu\bar \nu$
are similar four-body decays,
we suggest a relation, given by
\begin{eqnarray}\label{RB}
R(|{\cal \bar A}|^2)\equiv
\frac{|\bar {\cal A}(B^-\to \Lambda\bar p \nu\bar \nu)|^2}{|\bar {\cal A}(B^-\to p\bar p e^-\bar \nu_e)|^2}
= 3R(\text{Const}^2) \frac{1/m_{p\bar p}^{12}}{1/m_{\Lambda\bar p}^{12}}\,,
\end{eqnarray}
where the factor 3 comes from the three neutrino flavors and $R(\text{Const}^2)=0.012$ is
due to the constants of their own Hamiltonian
and the form factor for $f_2(g_2)$.
When the invariant masses $m_{p\bar p}$ and  $m_{\Lambda\bar p}$ are close to
1.877 and 2.054 GeV$^2$ for $B^-\to p\bar p e^-\bar \nu_e$ and $B^-\to \Lambda\bar p \nu\bar \nu$,
respectively, the curves are drawn to be at the top in the spectra.
Thus, we  obtain $R(|{\cal \bar A}|^2)\simeq \,R(\text{Const}^2)$,
which agrees with the numerical result
$R({\cal B})\equiv{\cal B}(B^-\to \Lambda\bar p \nu\bar \nu)/{\cal B}(B^-\to p\bar p e^-\bar \nu_e)=0.012$.
It is interesting to point out that
the measurement for $R({\cal B})$ can be a test of $1/t^3$ as the momentum dependence
of the $\bar B\to {\bf B\bar B'}$ transition form factors in Eqs. (\ref{transitionF}) and (\ref{figi}).

Due to the rich spin structure in the final state,  the baryonic decay of  $B^-\to \Lambda\bar p \nu\bar \nu$ is
clearly quite different from  the mesonic one of $B^-\to (K\pi)^- \nu\bar \nu$. The spin effect is sensitive to some new physics.
For example, the angular distributions in $B^-\to \Lambda\bar p \nu\bar \nu$
 can be used to probe for right-handed and (pseudo-)scalar currents beyond the SM.
In Ref.~\cite{IS}, the invisible scalar (S) decay has been studied for the mesonic decays
of $\bar B\to K^* SS$ and $\bar B\to K^{(*)}\nu\bar \nu$. 
Similar studies can be extended to the baryonic modes here.
In particular, we would like to emphasize that
to test the invisible scalar pair $SS$ from the $b\to sSS$ via the (pseudo-)scalar couplings,
 $B^-\to \Lambda\bar p\nu\bar \nu$ can be more beneficial than  $\bar B\to K^{(*)}\nu\bar \nu$.
As shown in Ref.~ \cite{Altmannshofer} that
the angular distributions in $\bar B\to K^*(\to K\pi)SS$ and $\bar B\to K^{(*)}(\to K\pi)\nu\bar \nu$
are  both angular-symmetric, one cannot distinguish them from 
 the angular analysis.
However, the situation for the baryonic decays are different.
Recall  that the large angular asymmetry observed to be $60\%$ in the $B^-\to p\bar p K^-$ decay~\cite{ppKLambdapbarpi_Belle}
has been attributed to the $\bar B\to {\bf B\bar B'}$ transition
via the (pseudo-)scalar couplings~\cite{GengHsiao3}.
Since the decay of $B^-\to \Lambda\bar p SS$ through $b\to sSS$
has the same type of the $\bar B\to {\bf B\bar B'}$ transition,
we expect it to be largely angular-asymmetric,
whereas  ${\cal A}_{\theta_B}(B^-\to \Lambda\bar p\nu\bar \nu)$
is predicted to be as small as 1~\%. 
If the integrated angular asymmetry in $B^-\to \Lambda\bar p SS$ is  50\%,  
to measure it at the $n\sigma$ level, about $5\times 10^{8}n\,B^\pm$
are required, which should be accessible to the future
B factories.

Finally, we remark that in  $B^-\to \Lambda \bar p\nu\bar \nu$,
as the spins and momentums may not be on the same plane,
similar to the cases in the $\bar B\to {\bf B\bar B'}M$ decays~\cite{GengHsiao6},
 $T$-odd triple product correlations (TPC's), such as
$\vec{p}_{\nu_e}\cdot (\vec{p}_\Lambda\times \vec{p}_{\bar p})$ and
$\vec{s}_\Lambda \cdot (\vec{p}_\Lambda\times \vec{p}_{\bar p})$
with $\vec{s}_\Lambda$ denoting the $\Lambda$ spin, can be generated. In the SM, since the decay depends on $\lambda_t=V_{ts}^*V_{tb}$,
which contains no CP phase, these T-odd observables are expected to be vanishingly small. However,
they can be used
to test direct $T$ violating effects from new particles.

\section{Conclusions}
We have studied the four-body semileptonic baryonic decay of $B^-\to \Lambda\bar p\nu\bar \nu$
based on the effective Hamiltonian for $b\to s\nu\bar \nu$,
arising from  electroweak penguin and box diagrams  in the SM.
We have calculated the decay branching ratio and angular distribution asymmetries for the decay.
Explicitly, we have found that ${\cal B}(B^-\to\Lambda\bar p\nu\bar\nu)=(7.9\pm 1.9)\times 10^{-7}$.
We have also obtained a useful relation between the decays of $B^-\to\Lambda\bar p\nu\bar\nu$
and $B^-\to p\bar p e^-\bar \nu_e$.
Similar to the rare mesonic decays of $B^-\to K^{(*)-}\nu\bar \nu$, the experimental search for
the rare baryonic decay of $B^-\to\Lambda\bar p\nu\bar\nu$ in the current as well as future $B$ factories
is useful to test the SM and limit new physics.

\begin{acknowledgments}
The work was supported in part by National Center of Theoretical Science
and  National Science Council (NSC-98-2112-M-007-008-MY3) of R.O.C.
\end{acknowledgments}


\begin{thebibliography}{99}
\bibitem{BexMnunu1}
B.~Aubert {\it et al.}  [BABAR Collaboration], Phys.\ Rev.\ Lett.\  {\bf 94}, 101801 (2005).
\bibitem{BexMnunu2}
B.~Aubert {\it et al.}  [BABAR Collaboration], Phys.\ Rev.\ Lett.\  {\bf 99}, 221801 (2007).
\bibitem{BthMnunu1}  
W.~Altmannshofer, A.~J.~Buras, D.~M.~Straub and M.~Wick, JHEP {\bf 0904}, 022 (2009).
\bibitem{ED}
P.~Colangelo, F.~De Fazio, R.~Ferrandes and T.~N.~Pham, Phys.\ Rev.\  D {\bf 73}, 115006 (2006).
\bibitem{UP}
T.~M.~Aliev, A.~S.~Cornell and N.~Gaur, JHEP {\bf 0707}, 072 (2007).
\bibitem{IS}
C.~S.~Kim, S.~C.~Park, K.~Wang and G.~Zhu, Phys.\ Rev.\  D {\bf 81}, 054004 (2010).
\bibitem{4th}
 A.~Arhrib and W.~S.~Hou,
  JHEP {\bf 0607}, 009 (2006).
\bibitem{ChenGeng}
C.~H.~Chen and C.~Q.~Geng, Nucl.\ Phys.\  B {\bf 636}, 338 (2002).
\bibitem{ppenu}
C.~Q.~Geng and Y.~K.~Hsiao, Phys.\ Lett.\  B {\bf 704}, 495 (2011)
\bibitem{ppKLambdapbarpi_Belle}
M.Z.~Wang {\it et al.} [BELLE Collaboration], Phys.\ Lett.\  B {\bf 617}, 141 (2005).
\bibitem{ppK(star)pi_Belle} M.Z. Wang {\it et al}. [BELLE Collaboration], Phys. Rev. Lett. {\bf 92}, 131801 (2004).
\bibitem{ppK_Babar}
B.~Aubert {\it et al.}  [BABAR Collaboration], Phys.\ Rev.\  D {\bf 72}, 051101 (2005).
\bibitem{ppKpi_Belle}
J.T.~Wei {\it et al.} [BELLE Collaboration], Phys.\ Lett.\  B {\bf 659}, 80 (2008).
\bibitem{pppi_Babar} 
B.~Aubert {\it et al.}  [BABAR Collaboration], Phys.\ Rev.\  D {\bf 76}, 092004 (2007).
\bibitem{ppKstar_Belle} 
J.H.~Chen {\it et al.}  [BELLE Collaboration], Phys.\ Rev.\ Lett.\  {\bf 100}, 251801 (2008).
\bibitem{ppD(star)_Belle} 
K.~Abe {\it et al.}, [BELLE Collaboration], Phys. Rev. Lett. {\bf 89}, 151802 (2002).
\bibitem{ppD(star)_Babar} 
B.~Aubert {\it et al.}  [BABAR Collaboration], Phys.\ Rev.\  D {\bf 74}, 051101 (2006).
\bibitem{Hou2}
C.K.~Chua, W.S.~Hou and S.~Y.~Tsai, Phys.\ Rev. D {\bf 66}, 054004 (2002).
\bibitem{GengHsiao3} 
C.Q.~Geng and Y.K.~Hsiao, Phys.\ Rev.\  D {\bf 74}, 094023 (2006).
\bibitem{Hsiao}
Y.K.~Hsiao, Int.\ J.\ Mod.\ Phys.\  A {\bf 24}, 3638 (2009).
\bibitem{GengHsiao5} 
C.Q.~Geng, Y.K.~Hsiao and J.~N.~Ng, Phys.\ Rev.\  D {\bf 75}, 094013 (2007).
\bibitem{GengHsiao2}
C.Q.~Geng and Y.K.~Hsiao, Phys. Lett. B {\bf 619}, 305 (2005).
\bibitem{GengHsiaoHY}   
C.H.~Chen, H.Y.~Cheng, C.Q.~Geng and Y.~K.~Hsiao, Phys.\ Rev.\  D {\bf 78}, 054016 (2008).
\bibitem{GengHsiao4}
C.Q.~Geng, Y.K.~Hsiao and J.~N.~Ng, Phys.\ Rev.\ Lett.\  {\bf 98}, 011801 (2007).

\bibitem{LambdaLambdaD} 
Y.~W.~Chang {\it et al.}  [BELLE Collaboration], Phys.\ Rev.\  D {\bf 79}, 052006 (2009).
\bibitem{LambdapbarD}
 P.~Chen {\it et al.}  [BELLE Collaboration], Phys.\ Rev.\  D {\bf 84}, 071501 (2011).
\bibitem{ppKstar_BABAR}
B.~Aubert {\it et al.} [BABAR Collaboration], Phys.\ Rev.\  D {\bf 76}, 092004 (2007).
\bibitem{ppKstar_BELLE} 
J.~H.~Chen {\it et al.}  [BELLE Collaboration], Phys.\ Rev.\ Lett.\  {\bf 100}, 251801 (2008).
\bibitem{btosnunu} 
T.~Inami and C.~S.~Lim, Prog.\ Theor.\ Phys.\  {\bf 65}, 297 (1981) [Erratum-ibid.\  {\bf 65}, 1772 (1981)].
\bibitem{topfn1}
G.~Belanger and C.~Q.~Geng, Phys.\ Rev.\  D {\bf 43}, 140 (1991).
\bibitem{topfn2}
  G.~Buchalla and A.~J.~Buras, Nucl.\ Phys.\  B {\bf 400}, 225 (1993).

\bibitem{Brodsky1}
G.P.~Lepage and S.J.~Brodsky, Phys.\ Rev.\ Lett.\  {\bf 43}, 545(1979) [Erratum-ibid.\  {\bf 43}, 1625 (1979)].
\bibitem{Brodsky2}
G.P.~Lepage and S.J.~Brodsky, Phys.\ Rev.\ D {\bf 22}, 2157 (1980).
\bibitem{Brodsky3}
S.J.~Brodsky, G.P.~Lepage and S.~A.~A.~Zaidi, Phys.\ Rev.\ D {\bf 23}, 1152 (1981).

\bibitem{Kl4}
A. Pais and S.B. Treiman, Phys. Rev. {\bf 168}, 1858 (1968).
\bibitem{Wise}
C.L.Y.~Lee, M.~Lu and M.B.~Wise, Phys.\ Rev.\  D {\bf 46}, 5040 (1992).
\bibitem{Cheng:1993ah}
H.Y.~Cheng, C.Y.~Cheung, W.~Dimm, G.L.~Lin, Y.C.~Lin, T.M.~Yan and H.L.~Yu, Phys.\ Rev.\  D {\bf 48}, 3204 (1993).
\bibitem{pdg} K.~Nakamura {\it et al.}  [Particle Data Group], J.\ Phys.\ G {\bf 37}, 075021 (2010).

\bibitem{HY1}
H.Y.~Cheng and K.C.~Yang, Phys. Rev. D {\bf 66}, 014020 (2002).
\bibitem{HY2}
H.Y.~Cheng and K.C.~Yang, Phys. Rev. D {\bf 66}, 094009 (2002).
\bibitem{Hou1}
C.K.~Chua, W.S.~Hou and S.Y.~Tsai, Phys. Rev. D {\bf 65}, 034003 (2002).
\bibitem{Hou3}
C.K.~Chua and W.S.~Hou, Eur.\ Phys.\ J.\ {\bf C29}, 27 (2003).
\bibitem{GengHsiao1}
C.Q.~Geng and Y.K.~Hsiao, Phys.\ Rev.\  D {\bf 75}, 094005 (2007).
\bibitem{Altmannshofer}
W.~Altmannshofer, A.~J.~Buras, D.~M.~Straub and M.~Wick,   JHEP {\bf 0904}, 022 (2009).

\bibitem{GengHsiao6}
C.Q.~Geng and Y.K.~Hsiao,
Phys. Rev. D {\bf 72}, 037901 (2005);
Int.\ J.\ Mod.\ Phys.\  A {\bf 21}, 897 (2006);
Int.\ J.\ Mod.\ Phys.\  A {\bf 23}, 3290 (2008).




\end{thebibliography}
\end{document}